\documentclass[aps,twocolumn,floatfix]{revtex4}
\usepackage{graphicx}
\usepackage{epsfig}
\def \Hm {{\underline H}^{{\rm mix}}_{\sigma }}
\def \lo {\lambda }
\def \ld {{L_z^{cf}}}
\def \lc {l_c}
\def \tcomp {T_{\rm comp}}

\def \TC {T_{\rm c}}

\def \vS {{\bf S}}
\def \vL {{\bf L}}

\def \Ms {M_S}
\def \Mp {M_{S'}}
\def \Ml {M_L}

\def \ts {t_{\sigma }}
\def \ap {\alpha }
\def \gf {\gamma_3}
\def \gs {\gamma_4}
\def \sgn {{\rm sgn}}
\def \TJU {T_{\rm JT}^{(u)}}
\def \TJL {T_{\rm JT}^{(l)}}
\def \ZTW {Z_{{\rm II}}}
\def \ZTH {Z_{{\rm III}}}
\def \mavg {M^{{\rm avg}}}
\def \on {{\bf 1}}
\def \tw {{\bf 2}}
\def \th {{\bf 3}}
\begin{document}

\date{\today}
\title{Jahn-Teller Distortion in Bimetallic Oxalates}
\author{Randy S. Fishman, Satoshi Okamoto, and Fernando A. Reboredo}
\affiliation{Materials Science and Technology Division, Oak Ridge National Laboratory, 
Oak Ridge, Tennessee 37831-6065}

\begin{abstract}

A C$_3$-symmetric crystal-field potential in the Fe(II)Fe(III) bimetallic oxalates 
splits the $L=2$ Fe(II) multiplet into two doublets and a singlet.  In compounds that exhibit 
magnetic compensation, one of the doublets was predicted to lie lowest in energy and carry 
a non-quenched orbital angular momentum $\pm \ld $, where $\ld $ exceeds a threshold value.  
In a range of $\ld $, a Jahn-Teller (JT) distortion increases the energy splitting
of the low-lying doublet and breaks the C$_3$ symmetry of the bimetallic planes around the 
ferrimagnetic transition temperature.  At low temperatures, the JT distortion 
disappears in compounds that display magnetic compensation
due to the competition with the spin-orbit coupling.  A comparison with recent measurements
provides strong evidence for this re-entrant, low-temperature JT transition
and a prediction for the normal, high-temperature JT transition.
The size of the JT distortion is estimated using first-principles calculations, which suggest that the 
long-range ordering of smaller, non-C$_3$-symmetric organic cations can eliminate 
magnetic compensation. 

\end{abstract}
\maketitle

\newpage

\section{Introduction}

Jahn-Teller (JT) transitions \cite{JT} in which electronic degeneracies are removed
by crystal distortions have been observed quite frequently in recent studies of
molecule-based magnets \cite{Jmol}.  It is well-known that the JT transition will be 
quenched if the degenerate levels carry orbital angular momentum and the 
spin-orbit coupling is sufficiently strong.  
When the spin-orbit coupling and JT energies are comparable, however,
a pseudo-JT transition with rather interesting behavior is possible.  In this paper,
we study the JT transition in a class of molecule-based magnets where the
spin-orbit coupling can be modified by choosing different organic cations to lie
between bimetallic layers.  Due to the competition between the spin-orbit
and JT energies, the JT distortion may vanish at low temperatures in a re-entrant,
first-order transition.

One of the most fascinating classes of molecule-based magnets,
bimetallic oxalates A[M(II)M'(III)(ox)$_3$] were first
synthesized \cite{Tamaki92} in 1992.  In the open honeycomb structure
of each bimetallic layer, sketched in Fig.1(a), the transition-metal ions M(II) and M'(III) are
coupled by the oxalate bridges ox=C$_2$O$_4$ \cite{Clem03}.  Depending on the
metal atoms, a single bimetallic layer can be either ferromagnetic or
ferrimagnetic (M(II) and M'(III) moments parallel or anti-parallel)
with magnetic moments pointing out of the plane.  While the organic
cation A separating the layers cannot alter the sign of the exchange
coupling, it does affect the overall properties of the system.

Some Fe(II)Fe(III) bimetallic oxalates exhibit magnetic compensation (MC)
due to cancellation of the moments on the Fe(II) and Fe(III) sublattices 
below the ferrimagnetic transition temperature $\TC $ \cite{comp}.
Based on symmetry and energy considerations, we recently explained why  
MC occurs for some organic cations but not for others \cite{Fish07}.  
The C$_3$-symmetric crystal field (invariant under in-plane 
rotations of $2\pi /3$) splits the $L=2$ Fe(II) multiplet into two doublets and a singlet.  
By shifting the Fe(II) ions with respect to the oxalate molecules, the cation A
determines the average orbital angular momentum $\pm \ld $ carried by the low-energy doublet.  
Compounds exhibit MC when the low-energy doublet lies 
below the singlet and $\ld $ exceeds the threshold $\lc $.  
For compounds that do not exhibit MC (``normal'' compounds), either $\ld < \lc $
or the singlet state lies lowest in energy.  

In the absence of spin-orbit coupling or when $\ld =0$, a JT distortion that breaks 
C$_3$ symmetry will always split the low-energy doublet. 
For large enough $\ld $, the spin-orbit coupling will quench the JT distortion.  
However, in a range of $\ld $ that includes $\lc $, we obtain a JT-distorted phase between 
temperatures $\TJL $ and $\TJU $ that bracket $\TC $.  
Both the re-entrant, low-temperature transition $\TJL $ and the normal,
high-temperature transition $\TJU $ are first order.  
Comparison with recent experiments \cite{Tang07} allows us to estimate
the normal JT transition temperature $\TJU $ in MC compounds.

This paper is divided into 5 sections.  Section II develops the phenomenological model
for the JT transition, with results described in Section III.  First-principles calculations
are described in Section IV and a discussion is contained in Section V.

\section{Model for the Jahn-Teller Transition}

Our model assumes a hierarchy of three energy scales.  The dominant energy is 
the Hund's coupling that determines the
spins $S=2$ and $S'=5/2$ on the Fe(II) (3d$^6$) and Fe(III) (3d$^5$) sites.
Next in importance is the C$_3$-symmetric crystal field $V$ generated by
the 6 oxygen atoms around each of the Fe sites.  These 6 oxygen atoms form
two triangles of slightly different sizes rotated by about 48 degrees with
respect to each other, one above and the other below the plane of Fe atoms.  
The smallest energies are the antiferromagnetic 
exchange coupling $J_c \vS \cdot \vS'$ between the Fe(II) and Fe(III) moments within
each bimetallic layer, 
the spin-orbit coupling $\lo \vL \cdot \vS $ on the Fe(II) sites ($\lo \approx -12.65$ 
meV \cite{Bleaney53}), and any non-C$_3$-symmetric contributions of the crystal potential.

As shown in Fig.1(b), the C$_3$-symmetric potential $V$ splits
the $L=2$ multiplet of the Fe(II) sites into two doublets $\psi_{1,2} $
and $\psi_{4,5} $, and a singlet $\psi_3  $.  The low-energy
doublet $\psi_{1,2} $ carries an average orbital angular momentum $\pm \ld $,
where $\ld $ ranges from 0 to 2 depending on the crystal-field parameters.
Of course, the singlet state $\psi_3 $ does not carry any orbital angular
momentum.  

\begin{figure}
\includegraphics *[scale=0.3]{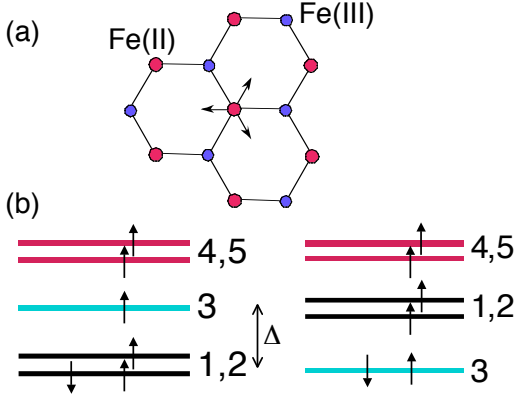}
\caption{
(Color online) (a) A portion of the open honeycomb lattice, displaying
three equivalent displacements of the Fe(II) ion into the adjacent hexagons.
(b) Two possible level-splitting schemes of the 3d$^6$ Fe(II) orbitals in 
a C$_3$-symmetric crystal-field, with either
the doublet $\psi_{1,2}$ or the singlet $\psi_3$ lowest in energy.  }
\end{figure}

Including the orbital contribution, the Fe(II) magnetic moment 
can be written $M(T)=\Ms(T)+\Ml(T) \le  0$, where $\Ms=2\langle S_z\rangle $
and $\Ml = \langle L_z\rangle $.  Of course, the Fe(III) magnetic moment 
$M'(T)=\Mp(T)=2\langle S'_z\rangle \ge 0$ has no orbital contribution.  
The mean-field (MF) approximation is used to treat the nearest-neighbor exchange 
\begin{equation}
J_c\vS \cdot \vS'\approx J_c \Bigl\{ \bigl(\Ms S'_z + S_z \Mp \bigr)/2 - \Ms \Mp /4  \Bigr\}
\end{equation} 
between neighboring Fe(II) and F(III) spins.  The MF eigenvalues of $\psi_{1,2;\sigma }$
and $\psi_{3\sigma }$ are
\begin{equation}
\label{ep1}
\epsilon_{1\sigma }=(-\vert \lo \vert \ld +3J_c \Mp /2)\sigma,
\end{equation}
\begin{equation}
\label{ep2}
\epsilon_{2\sigma }=(\vert \lo \vert \ld +3J_c \Mp /2)\sigma ,
\end{equation} 
\begin{equation}
\epsilon_{3\sigma }=(3J_c \Mp  /2)\sigma + \Delta ,
\end{equation}
so that the doublet $\psi_{1,2;\sigma }$ is split by the spin-orbit coupling
with $\vert \epsilon_{2\sigma }-\epsilon_{1\sigma }\vert =2\vert \lo \sigma \vert \ld $.
Because the crystal-field potential is assumed much larger than the spin-orbit
and exchange energies, $\vert \Delta \vert \gg J_c$ or $\vert \lo \vert $.

Using Eqs.(\ref{ep1}) and (\ref{ep2}) for the eigenvalues $\epsilon_{1,2;\sigma }$, it is 
straightforward to evaluate $M(T)$, $M'(T)$, and the average magnetization 
$\mavg (T)=(M(T)+M'(T))/2=(\vert M'(T)\vert -\vert M(T)\vert )/2$.
When $\lc \approx 0.238 < \ld < 1$, $\mavg (T)$ passes through 0 at the compensation temperature
$\tcomp < \TC $.  Just below $\TC $, $\mavg (T) < 0$ because the parallel
spin-orbit coupling between $\vL $ and $\vS $ causes the Fe(II) moment
$M(T)$ to increase in amplitude more rapidly
than the Fe(III) moment $M'(T)$.  But at zero temperature, the Fe(III) 
moment saturates at the larger value 
$\vert M'(0)\vert  = 2S' > \vert M(0)\vert  = 2S + \ld $ so that $\mavg (0) > 0$.

After comparing the observed values of $\tcomp /\TC \approx 0.62$ and $\TC \approx 45$ K
with the theoretical predictions, we estimated \cite{Fish07}
that $J_c\approx 0.46$ meV and $\ld \approx 0.28$
in MC compounds.  Normal compounds  
can fall into two categories.  Either the doublet remains lower in
energy than the singlet ($\Delta > 0$) but with $\ld < \lc'  \approx 0.234$ or the singlet lies lowest 
in energy ($\Delta < 0$).  Our model also predicted that two compensation points were possible 
in the narrow window $\lc'  < \ld < \lc $.  The recent observation by 
Tang {\em et al.} \cite{Tang07} of two compensation points in the compound 
N(n-C$_4$H$_9$)$_4$[Fe(II)Fe(III)ox$_3$] 
(data shown in Fig.3) would seem to confirm this prediction.

If a doubly-degenerate level is occupied by a single electron in the absence of
spin-orbit coupling, then a local JT displacement the 
Fe(II) ions corresponding to one of three equivalent directions is
always favored by the electronic energy.  The eigenstates
$\psi_{1\sigma } $ and $\psi_{2\sigma } $ on the Fe(II) sites are
mixed by the JT distortion as described by the Hamiltonian \cite{mix}
\begin{equation}
 \Hm  = \left( \begin{array}{cc}
\epsilon_{1\sigma } & \xi   \\
\xi & \epsilon_{2\sigma } \\
\end{array} \right),
\end{equation}
where $\xi $ is independent of $\sigma $.  
If $\epsilon_{1\sigma }=\epsilon_{2\sigma }$ are given by the MF result
$\epsilon_{0\sigma }\equiv 3J_c M'\sigma /2$, 
then the eigenstates of $\Hm $ are $\psi_{a\sigma }=(1/\sqrt{2})(\psi_{1\sigma }
+\psi_{2\sigma })$
and $\psi_{b\sigma }=(1/\sqrt{2})(\psi_{1\sigma }-\psi_{2\sigma })$, with 
$L_{a,b;\sigma } =\langle \psi_{a,b;\sigma } \vert L_z \vert \psi_{a,b;\sigma }\rangle = 0$ and 
eigenvalues $\epsilon_{a,b; \sigma }=\epsilon_{0\sigma } \pm \xi $.  
So in the absence of spin-orbit coupling, the orbital angular momentum is quenched by the 
JT distortion and the doublet $\psi_{a,b;\sigma }$ is split by $2\vert \xi \vert $.  
The $T=0$ energy is then given by $E/N=-3J_cSS'-\xi + \ap \xi^2 /\vert \lo \vert $, 
where the second term is an elastic restoring potential and $N$ is the number of Fe(II) or
Fe(III) sites per bimetallic layer.  Since $\ap > 0$, the $T=0$ equilibrium value
for $\xi $ is $\vert \lo \vert /2\ap > 0$.

Including the spin-orbit interaction $\lo \vL \cdot \vS$, the eigenvalues of $\Hm $ are  
$\epsilon_{a\sigma } = \epsilon_{0\sigma } + \ts $ and 
$\epsilon_{b\sigma } =\epsilon_{0\sigma }-\ts $,
where $\ts = -\sgn (\sigma ) \sqrt{ (\lo \ld \sigma)^2+\xi^2}$.  Hence, the
doublet splitting $\delta \equiv 2\vert \ts \vert $ is enhanced by the JT
effect (strictly speaking, the pseudo-JT effect when 
$\epsilon_{1\sigma }\ne \epsilon_{2\sigma }$).  
Since there is no spin-orbit coupling when $\sigma =0 $, 
$t_0=\xi $ and $L_{a0}=L_{b0}=0$.  For $\sigma \ne 0$, the 
spin-orbit coupling maintains a nonzero orbital angular momentum 
$L_{a\sigma }=-L_{b\sigma }$ even in the presence of a JT distortion:
\begin{equation}
\label{La}
L_{a\sigma }  =
(\ld )^2 \lo \sigma \frac{\lo \ld \sigma +\ts }{(\lo \ld \sigma)^2
+\lo \ld \sigma \ts+\xi^2},
\end{equation}
so that $L_{a\sigma }=L_{a,-\sigma}$.
When $\xi \ne 0$, the JT distortion suppresses the absolute values
$\vert L_{a\sigma }\vert < \ld $ for $\sigma =\pm 1$ and $\pm 2$.

With a correction to avoid double counting, the MF free energy can be written 
\begin{eqnarray}
\label{Fr}
&&\frac{F}{N}=-T\log \Bigl\{ \ZTW \ZTH e^{3J_c \Ms \Mp /4T} \Bigr\} \nonumber \\ 
&&+\ap \vert \lo \vert \biggl\{
\biggl( \frac{\xi }{\vert \lo \vert } \biggr)^2 +\gf \biggl( \frac{\xi }{\vert \lo \vert }\biggr)^3 
+\gs \biggl( \frac{\xi }{\vert \lo \vert }\biggr)^4
\biggr\}, \
\end{eqnarray}
where 
\begin{equation}
\ZTW =2\sum_{\sigma} e^{-3J_c \Mp \sigma /2T} \cosh (\ts /T),
\end{equation}
\begin{equation}
\ZTH =2\sum_{\sigma'} e^{-3J_c \Ms \sigma' /2T}.
\end{equation}
The sums in the partition functions $\ZTW $ and $\ZTH $ are over 
$\sigma = 0,\, \pm 1,\, \pm 2$ and $\sigma' = \pm 1/2,\, \pm 3/2,\, \pm 5/2$.
The second line in Eq.(\ref{Fr}) corresponds to the elastic energy.

Breaking C$_3$ symmetry, a local JT distortion with $\pm \xi $ involves
the displacement of an Fe(II) atom either into one of the three open hexagons or
towards one of the three neighboring Fe(III) atoms.  The former distortions
are sketched in Fig.1(a).  The anharmonic $\gf (\xi /\vert \lo \vert)^3$ term in Eq.(\ref{Fr}) reflects
the different energy costs for those two types of distortions.
So the anharmonicity and source of the first-order JT transition arise 
quite naturally on the open honeycomb lattice.  The sign of $\gf $ does not affect
any physical results and is chosen to be negative only so that $\xi \ge 0$.
Any fluctuations between the distorted atomic configurations are assumed slow 
compared to the electronic time scales.  

It is simple to obtain the equilibrium values for $\Ms $ and $\Mp $ from the
extremal conditions $\partial F/\partial \Ms =\partial F/\partial \Mp =0$.   Because  
the spin-orbit energy $\lo \vL \cdot \vS$ is treated exactly, $\Ml $
is not a variational parameter in the MF free energy and must be determined separately 
from the condition 
\begin{equation}
\Ml = -\frac{2}{\ZTW }\sum_{\sigma } L_{a\sigma } e^{-3J_c \Mp \sigma /2T} \sinh (\ts /T).
\end{equation}
Of course, the equilibrium value for the JT parameter $\xi $ is obtained by minimizing
$F$ with respect to $\xi $.

\section{Model Results}

The average magnetization and JT energy $\xi $ are plotted versus temperature 
in Figs.2(a) and (b) for $\ap =3.7$, $\gf =-1.9$, and $\gs =1.1$.  
As expected, the MC threshold $\lc \approx 0.282$ is enhanced by 
the JT distortion.  When $\ld < 0.260$, the JT distortion $\xi (T) $
persists down to $T=0$.  For $\ld =0$ (but
still taking $\Delta > 0$), $\xi (0) \approx 0.842\vert \lo \vert \approx 10.5$ meV and
$\TJU \approx 0.578 \vert \lo \vert $.  For $\ld =0.2 $, the JT distortion would increase
the splitting of the $\sigma = 2$ doublet from $4\vert \lo \vert \ld = 10$ meV
to $2\sqrt{ (2\lo \ld)^2+\xi^2} \approx  23$ meV.  

When $\ld \ge 0.260$, the JT distortion is quenched at $T=0$ due to the 
strong orbital ordering.  Hence, we obtain both lower and upper JT transitions, 
$\TJL $ and $\TJU $.   At the re-entrant JT transition, the distortion $\xi $ vanishes below 
$\TJL $ and appears above $\TJL $.
The temperature range $\TJU - \TJL $ decreases as $\ld $ increases and vanishes when
$\ld > 0.324$.  Notice that the JT transitions at $\TJU $ and $\TJL $
always bracket $\TC $.  

\begin{figure}
\includegraphics *[scale=0.6]{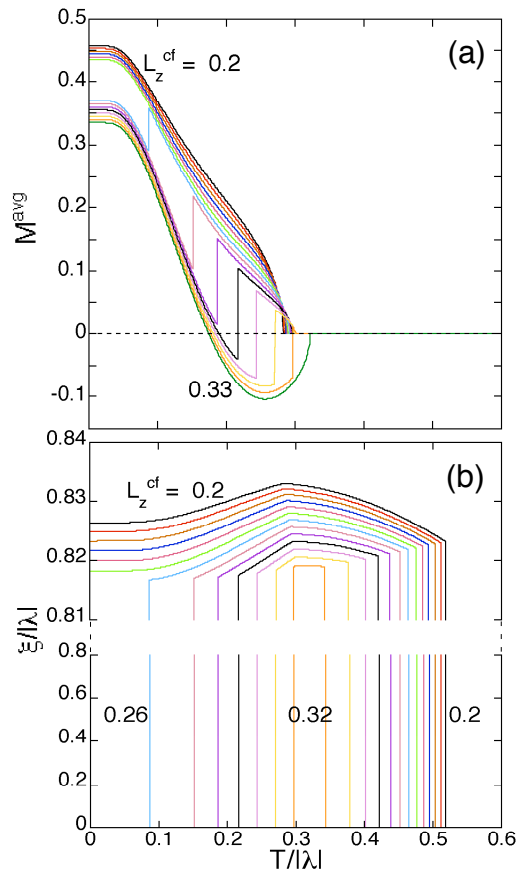}
\caption{
(Color online) The (a) average magnetization and (b) the JT mixing energy $\xi $ 
normalized by $\vert \lo \vert $ versus
temperature $T/\vert \lo \vert $ for a range of $\ld $ in increments of 0.01
using $J_c/\vert \lo \vert  = 0.037$ and the elastic constants given in the text.
}
\end{figure}

Because the orbital contribution $\Ml (T)$ to the Fe(II) moment drops as $\xi $
jumps at the re-entrant JT transition, $\TJL $ is marked by a discontinuous 
change in $\mavg (T)$.  With decreasing $\ld $, both the orbital contribution
$\Ml (T)$ and the magnitude of the change in $\mavg (T)$ become
smaller.  For $\lc < \ld \le 0.324$, the re-entrant JT transition causes the average 
magnetization to change sign, as shown in Fig.2(a) for $\ld =0.30$. 

\begin{figure}
\includegraphics *[scale=0.35]{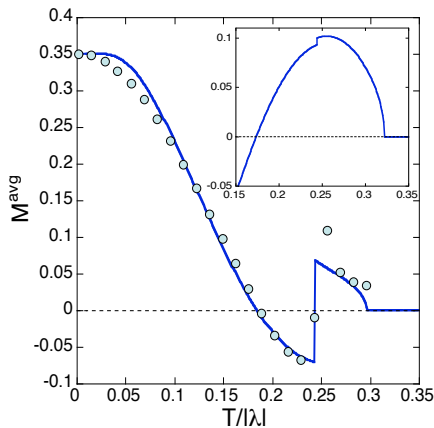}
\caption{
(Color online) The average magnetization versus $T/\vert \lo \vert $
for $\ld =0.30$ and the experimental data of Ref.\cite{Tang07} (rescaled so
that $\mavg (0)=0.35$)
using $J_c/\vert \lo \vert  = 0.037$ and the elastic constants given in the text.
Inset is the predicted average magnetization versus $T/\vert \lo \vert $ for 
a mixture of $\ld = 0.30$ (5\%) and $\ld =0.33$ (95\%) components, with both 
magnetizations chosen to be positive just below $\TC $.
}
\end{figure}

The quantitative agreement in Fig.3 between the theoretical prediction for 
$\ld =0.30$ and the measurements of Ref.\cite{Tang07} for A=N(n-C$_4$H$_9$)$_4$ 
is striking.  Rather than providing evidence for two compensation
points, Ref.\cite{Tang07} indicates that a re-entrant
JT transition occurs at $\TJL \approx 42$ K.  According to Fig.2(b), a normal
JT transition will be found at $\TJU \approx 70$ K.
Indeed, recent x-ray measurements \cite{Watts05} on the same compound
confirm that the hexagonal symmetry present
at room temperature is absent in the monoclinic lattice below 60 K.
 
Measurements on several Fe(II)Fe(III) compounds \cite{Watts05, Nut98}
suggest that all MC compounds exhibit a small jump in the
magnetization between $\TC $ and $\tcomp $.  The predicted jumps in
Fig.2(a) are much too large to explain those measurements.  X-ray
diffraction studies \cite{stf} reveal that stacking faults in several
compounds promote the coexistence of two phases: one with a six-layer
repeat and the other with a two-layer repeat.  The small magnetization
jumps observed \cite{Watts05, Nut98} in MC compounds are likely caused
by a mixture of those two stacking types.  As shown in the inset to
Fig.3, a mixture of two phases, type 1 with $\ld = 0.30 $ (5\% of the
sample) and type 2 with $\ld = 0.33$ (95\% of the sample) \cite{mon},
produces a small jump which is quite similar to the observations.  The
much larger jump observed by Tang {\em et al.} \cite{Tang07} may be
caused by the greater fraction of type 1 ($\ld \approx 0.30$) stacking
in their sample.

While there is no JT distortion ($\xi =0$) when $\Delta < 0$,
$\xi (T)$ remains nonzero down to $T=0$ when $\Delta > 0$ and $\ld < 0.260$.
So our model predicts that normal bimetallic oxalates with $\ld < 0.260$
will not exhibit a discontinuity in the magnetization, which may explain
why such a jump has never been observed in a normal compound \cite{Nut98}.  
Nevertheless, normal Fe(II)Fe(III) compounds with $\Delta > 0$ (so that the doublet 
remains lowest in energy) should manifest a normal JT transition at 
$\TJU \approx 0.58 \vert \lo \vert $ or about 85 K. 

But any non-C$_3$-symmetric cation like N(n-C$_4$H$_9$)$_4$
will also induce a permanent distortion of the hexagonal lattice.  
Depending on the size and shape of the cation,
this distortion can be local, weakly-correlated, or long-ranged.
A non-C$_3$-symmetric potential can be included within our model by 
changing the off-diagonal terms in $\Hm $ from $\xi $ to $\xi +\xi_0$.
In the absence of spin-orbit coupling and a spontaneous JT distortion $\xi $,
the doublet splitting $\delta $ is then $2\xi_0$.  Since the anharmonic 
elastic potential proportional to $\xi^3$ favors $\xi > 0$, it acts to enhance the 
total distortion $\vert \xi + \xi_0 \vert $ when $\xi_0 >0$ and to suppress the total 
distortion when $\xi_0 < 0$.  For any nonzero $\xi_0$, there is a spontaneous JT 
distortion $\xi (T)\ne 0$ at all temperatures due to the linear
term of order $-\xi \xi_0 /\vert \lo \vert $ in the free energy.   

\begin{figure}
\includegraphics *[scale=0.5]{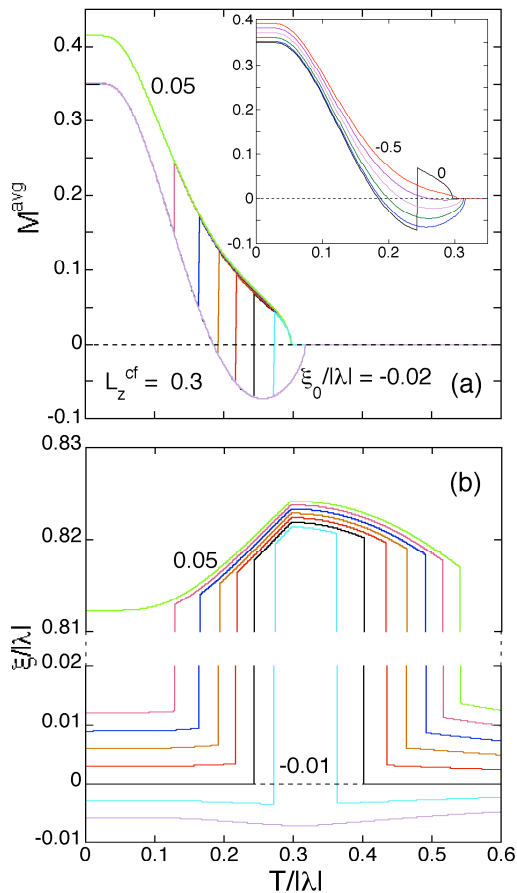}
\caption{
(Color online) The (a) average magnetization and (b) the JT mixing energy $\xi $ 
normalized by $\vert \lo \vert $ versus temperature $T/\vert \lo \vert $ for several 
values of the permanent distortion $\xi_0/\vert \lo \vert $ from -0.02 to 0.05 in increments
of 0.01.  Inset in (a) is the magnetization versus $T/\vert \lo \vert $ for $\xi_0/\vert \lo \vert $
between -0.5 and 0 in increments of 0.1.  These plots use 
$\ld =0.3$, $J_c/\vert \lo \vert  = 0.037$ and the elastic constants given in the text.
}
\end{figure}

As shown in Fig.4 for $\ld =0.3$ and the elastic parameters given earlier, increasing 
$\xi_0 $ for a fixed $\ld $ decreases $\TJL$ while keeping $\tcomp $
almost unchanged until $\TJL < \tcomp $.  When $\xi_0 > 0.047\vert \lo \vert \approx 0.6$ meV, 
the spontaneous JT distortion $\xi $ at $T=0$ jumps to a value near
$0.81\vert \lo \vert \approx 10$ meV and the lower JT transition disappears. 
When $\xi_0 < -0.019 \vert \lo \vert \approx -0.2$ meV, the spontaneous JT transition is 
eliminated but magnetic compensation survives until $\xi_0 < -0.43 \vert \lo \vert \approx -5.4$ meV,
as seen in the inset to Fig.4(a).  Except in the narrow window 
$0 > \xi_0 > -0.019 \vert \lo \vert $, the permanent
distortion $\xi_0$ and the electronic distortion $\xi $ have the same sign.

Clearly, the electronic JT distortion $\xi $ will vanish if the Fe(II) singlet
lies below the doublet with $\Delta < 0$.
So two conditions are required for C$_3$ symmetry to remain unbroken at $T=0$.
First, the cation must preserve C$_3$ symmetry so that there is no permanent
distortion $\xi_0$.  Second, either $\Delta < 0 $ or if $\Delta > 0$, 
the average orbital angular momentum $\ld $ of the doublet must be
sufficiently large to quench the spontaneous JT distortion $\xi $ ($\ld \ge 0.260$ in Fig.2).  
Specific examples of symmetry breaking will be discussed in the next section.

\section{First-Principles Calculations}

In order to estimate the orders of magnitude of the spontaneous JT
distortion $\xi $ and the permanent distortion $\xi_0 $, 
we performed a series of first-principles calculations within the framework of
density-functional theory (DFT).  We 
employed the local spin density
approximation (SDA) in the plane-wave-pseudopotential approach with
the PBE \cite{PBE} exchange correlation functional as implemented in
the Quantum-ESPRESSO package \cite{pwscf}.  We used Vanderbilt ultrasoft
pseudopotentials \cite{pwscf,vanderbilt,rabe} including, in the case of
Fe \cite{pwscf,rabe},  d electrons in the valence and 
non-linear core corrections.  An energy cut-off of 45 Ry was used.
Since the orbitals at the Fermi level are very localized in this ionic solid, we
used a single {\bf k}-point.  Spin orbit coupling was not included.
In order to stabilize the electronic density, the calculations were
performed with an electronic temperature of 0.02 Ry or
270 meV.

Predictions of the experimental electronic and magnetic structure in
highly localized systems are known to be significant challenges for
most approximations of DFT.  Indeed, we find that the energy
difference between ferromagnetic and antiferromagnetic configurations
is below the resolution of our theory.  Moreover, the charge-density
wave observed experimentally (which results in inter-penetrated Fe(II)
and Fe(III) networks) is not found to be the ground state. This
possibly signals the presence of strong self-interaction errors for
the localized d orbitals \cite{perdew81}.  The antiferromagnetic
Fe(II) and Fe(III) ordering is stabilized only after enforcing the net
spin within each unit cell to be equal to 1/2.

Calculations were performed for several possible stackings ($ab$, $aab$,
$abc$) of the bimetallic Fe(II)-Fe(III) layers with A=N(n-C$_3$H$_7$)$_4$
cations.  The total energy differences between these
stackings were below the accuracy of the
theory, consistent with the high degree of
polymorphism and the low energy cost for stacking faults suggested by
the x-ray scattering experiments \cite{stf}.  Since the
relative energy difference $\Delta$ between the doublet and singlet is
quite sensitive to the stacking of the bimetallic planes, the angular
momentum $\ld $ of the ground state may also depend on the stacking of the
bimetallic planes.

The doublet splittings and the order of magnitude of the JT
distortions were estimated by focusing on a single bimetallic layer
with different cations A.  Within hexagonal supercells, we studied the
cations A=NH$_4$ ($\on $) and N(n-C$_3$H$_7$)$_4$ ($\tw $).  We also
considered A=N(n-C$_3$H$_7$)$_4$ ($\th $) in a non-hexagonal
supercell.

($\on $) Although no oxalates contain the cation
NH$_4$ because it is too small to stabilize the open honeycomb lattice,
calculations on this system allow us to estimate the JT
distortion in a C$_3$-symmetric environment.  First, we relaxed the
forces on all atoms constraining the symmetry to remain C$_3$.  We
found that the Fermi level for the majority band lies at a doublet
localized at the Fe(II) sites and occupied by a single electron.  This
configuration is consistent with our model, which requires the partial
occupation of a doublet to explain the MC and the JT distortion.  Second,
we displaced the Fe(II) ion from the symmetric position on a grid of
points ${\bf r}$.  From the {\it ab-initio} calculations, we extracted
the total energy $E({\bf r})$ of the configuration and the energy splitting
$\delta ({\bf r})$ induced in the partially-occupied doublet.  Because
the calculations where performed at an electronic temperature 
($\sim 270$ meV) much
larger than the splittings, both electronic levels of
the doublet have an equal occupation of $1/2$.  At this large
electronic temperature, no JT electronic energy was gained and the
symmetric point ${\bf r}=0$ remained the position with minimum energy, as expected.

We can, however, estimate the energy gain
at $T = 0$ as $E({\bf r})-\delta({\bf r})/2$.  This approximation assumes that the
total energy difference is given by the sum of the occupied
eigenvalues at zero temperature and neglects a small change in the
electronic density \cite{martin}.  With this expression, we obtain the
approximate Fe(II) position and doublet splitting at zero temperature. 
In absence of the spin-orbit coupling, we estimate that
the Fe(II) moves $0.03\, \AA $ (the nearest-neighbor Fe(II)-Fe(III) 
distance is about $5.4\, \AA $) with an energy gain of 2 meV and
a doublet spitting of 8 meV.  This JT splitting is quite close to
the spin-orbit coupling $\vert \lo \vert \ld S \approx 7.5$ meV with $\ld =0.3$.

($\tw $) Moderately-sized cations such as N(n-C$_3$H$_7$)$_4$ are themselves
non-C$_3$-symmetric and break the C$_3$ symmetry of the crystal-field
potential at the Fe(II) sites.  While it is possible to construct a
N(n-C$_3$H$_7$)$_4$ isomer with three C$_3$H$_7$ radicals below the
oxalate plane, thereby preserving $C_3$ symmetry, the remaining propyl
chain (oriented towards the hexagonal hole in the oxalates plane) must
break C$_3$ symmetry.  This symmetry violation produces a
crystal-field splitting $2\xi_0$ of the doublet and a permanent
distortion of the open honeycomb lattice.

Calculations were performed in a hexagonal unit-cell containing a
single A=N(n-C$_3$H$_7$)$_4$ cation with periodic boundary conditions,
corresponding to to an ordered configuration where every
N(n-C$_3$H$_7$)$_4$ cation is oriented in the same way.  
After relaxing the positions of the atoms, we
obtained an intrinsic doublet splitting $2\xi_0$ of about 10 meV at an
electronic temperature of 270 meV.

These calculations also indicate that the intrinsic distortion
introduced by the cation will be increased at low temperature 
by the electronic energy gain of the Fe(II) JT distortion.
So in the absence of spin-orbit coupling, the parameters $\xi $ and $\xi_0$ of 
our model have the same sign for this cation.  

($\th $) The molecule N(n-C$_3$H$_7$)$_4$ is just small enough to
allow individual cations to rotate independently of each other.
Hence, a uniform distortion of the crystal is possible.  Such a
uniform distortion of the open honeycomb structure has been observed in
bimetallic oxalates with radical cations \cite{nonc}.  For an
ordered arrangement of A=N(n-C$_3$H$_7$)$_4$ cations and allowing 
the supercell to break hexagonal symmetry, we
obtain a doublet splitting $\delta \sim 20$ meV, which should increase to
about 30 meV when the electronic temperature approaches zero.

By contrast, larger cations such as A=N(n-C$_4$H$_9$)$_4$ studied in
Ref.\cite{Tang07} are unable to independently rotate within each unit
cell.  Because they are locked into a highly-disordered configuration
during synthesis, such cations will only break the local C$_3$
symmetry around each Fe(II) ion.  Since the spin-orbit coupling
competes with the small, local lattice distortions, magnetic
compensation is still possible when the cations are large and
disordered.  This leads to an interesting conjecture: due to their
ability to spatially order, smaller cations like N(n-C$_3$H$_7$)$_4$
may be more effective at enhancing the departure from C$_3$ symmetry
and suppressing magnetic compensation than larger cations like
N(n-C$_4$H$_9$)$_4$.

\section{Discussion}

Our first-principles calculations imply that large cations like
N(n-C$_4$H$_9$)$_4$ will induce a substantial, local distortion $\xi_0
$ of the hexagonal lattice.  For smaller cations like
N(n-C$_3$H$_7$)$_4$, this distortion may propagate throughout the
bimetallic layer because the cations are able to reach a global
minimum of the free energy, which is an ordered state at low
temperatures.  For an ordered configuration of
N(n-C$_3$H$_7$)$_4$ cations with $\vert \xi_0 \vert \approx 5$ meV,
the value suggested by our first-principles results,
Fig.4 implies that the re-entrant JT transition will be absent and any
magnetic compensation will be very weak.  Indeed, 
N(n-C$_3$H$_7$)$_4$[Fe(II)Fe(III)ox$_3$] bimetallic oxalates
\cite{comp} show no signs of a JT transition or magnetic compensation
below $\TC $.

With the spin-orbit coupling set to zero ($\ld =0$), the doublet
splitting $\delta = 2\xi (0) \approx 21$ meV obtained using the parameters of
Fig.3 is more than twice as large as that obtained from
first-principles calculations in a C$_3$-symmetric environment (see
$\on $ above).  So it is likely that the organic cation
N(n-C$_4$H$_9$)$_4$ plays a significant role in breaking the local
C$_3$ symmetry and enhancing the doublet splitting.  A more
sophisticated description of the experimental measurements may be
possible once additional information about the atomic structure becomes available.

To summarize, we have provided strong evidence for the existence of a
re-entrant JT transition in the Fe(II)Fe(III) bimetallic oxalates.
Observation of the JT distortion between $\TJL \approx 42$ K and $\TJU
\approx 70$ K would provide unambiguous support for the predicted
multiplet splitting in MC compounds.  We hope that this work will
inspire systematic x-ray scattering measurements that will verify 
the predictions made in this paper, including the long-range
ordering of small, non-C$_3$-symmetric cations in this important
class of layered, molecule-based magnets.

We would like to acknowledge conversations with Dr. Murilo Tiago.
This research was sponsored by the Laboratory Directed
Research and Development Program of Oak Ridge National Laboratory,
managed by UT-Battelle, LLC for the U. S. Department of Energy
under Contract No. DE-AC05-00OR22725 and by the Division of Materials Science
and Engineering of the U.S. DOE.

\newpage


\begin{references}

\bibitem{JT} See for example, Bersuker, I. B. {\it The Jahn-Teller Effect};  Cambridge University Press:
Cambridge, 2006 and references therein.

\bibitem{Jmol} (a) Sorai, M.; Nakano, M.; Miyazaki, Y. {\it Chem. Rev.} {\bf 2006}, {\it 106}, 976;
(b) Beghidja, C.; Rogez, G.; Kortus, J.; Wesolek, M.; Welter, R. {\it J. Am. Chem. Soc.} {\bf 2006}, {\it 128},
3140; (c) Hatnean, J. A.; Raturi, R.; Lefebvre, J.; Leznoff, D. B.; Lawes, G.; Johnson, S. A. {\it J. Am. Chem. Soc.} 
{\bf 2006}, {\it 128}, 14992;  (d) Milios, C. J.; Vinslava, A.; Wernsdorfer, W.; Prescimone, A.; Wood, P. A.;
Parsons, S.; Perlepes, S. P.; Christou, G.; Brechin, E. K. {\it J. Am. Chem. Soc.} {\bf 2007}, {\it 129}, 6547.

\bibitem{Tamaki92} Tamaki, H.; Zhong, Z. J.; Matsumoto, N.; Kida, S.; Koikawa, M.; Achiwa, N.;
Hashimoto, Y.; \~Okawa, H. {\it J. Am. Chem. Soc.} {\bf 1992}, {\it 114}, 6974.

\bibitem{Clem03} See the review Cl\'ement, R.;  Decurtins, S.; Gruselle, M.; Train, C. {\it Mon.
f\"ur Chem.} {\bf 2003}, {\it 134}, 117.

\bibitem{comp} (a) Mathoni\`ere, C.; Carling, S. G.; Day, P. {\it J. Chem. Soc., Chem. Commun.} 
{\bf 1994}, 1551; (b) Mathoni\`ere, C.; Nuttall, C. J.; Carling, S. G.; Day, P., {\it Inorg. Chem.} 
{\bf 1996}, {\it 35}, 1201; (c) Clemente-Le\'on, M.; Coronado, E.; G\'omez-Garc\'ia, C. J.; 
Soriano-Portillo, A. {\it Inorg. Chem.} {\bf 2006}, {\it 45}, 5653.

\bibitem{Fish07} Fishman, R. S; Reboredo, F. A. {\it Phys. Rev. Lett.} {\bf 2007},
{\it 99}, 217203.

\bibitem{Tang07} Tang, G.; He, Y.; Liang, F.; Li, S.; Huang, Y. {\it Physica B} {\bf 2007}, {\it 392}, 337.

\bibitem{Bleaney53} Bleaney, B.; Stevens, K. W. H. {\it Rep. Prog. Phys.} {\bf 1953}, {\it 16}, 108. 

\bibitem{mix} Diagonal terms in $\Hm $ involving $\xi $ can be neglected because
the doublet states are related by the time-reversal operations
$\psi_{1\sigma }=-K \psi_{2,-\sigma }$ and $\psi_{2\sigma }=-K \psi_{1,-\sigma }$, 
where $K$ is the time-reversal operator.
So for any spin-independent electrostatic potential $V=K^{-1}VK$, 
$\langle \psi_{1\sigma } \vert V \vert \psi_{1\sigma }\rangle =
\langle \psi_{2,-\sigma } \vert K^{-1} V K\vert \psi_{2,-\sigma }\rangle =
\langle \psi_{2\sigma } \vert V \vert \psi_{2\sigma }\rangle $, leading only to a shift
of the doublet with respect to the singlet.
  

\bibitem{Watts05} Watts, I. D.; Carling, S. G.; Day, P.; Visser, D. {\it J. Phys. Chem. Sol.} {\bf 2005},
{\it 66}, 932.

\bibitem{Nut98} Nuttall, C. J.; Day, P. {\it Chem. Mat.} {\bf 1998}, {\it 10}, 3050.

\bibitem{stf} (a) Nuttall, C. J.; Day, P. {\it J. Sol. St. Chem.} {\bf 1999}, {\it 147}, 3;  
(b) Ovanesyan, N. S.; Makhaev, V. D.; Aldoshin, S. M.; Gredin, P.; Boubekeur, K.;
Train, C.; Gruselle, M. {\it Dalton Trans.} {\bf 2005}, {\it 18}, 3101.

\bibitem{mon}  In a Mn(II)Fe(III) compound, Ref.\cite{stf}(a) reported 
a mixture of two stacking types with a faulting probability between 20 and 30\%.
In the Fe(II)Fe(III) family, MC compounds were found to be more monophasic.

\bibitem{PBE} Perdew, J. P.; Burke, K.; Ernzerhof, M. {\it Phys. Rev. Lett.}
{\bf 1996,} {\it 77}, 3865.

\bibitem{pwscf} Baroni, S.; Dal Corso, A.; deGironocli, S.; Giannozzi, P.
http://www.pwscf.org.

\bibitem{vanderbilt} Vanderbilt, D. {\it Phys. Rev. B} {\bf 1990}, {\it 41}, 7892.

\bibitem{rabe} Rappe, A. M.; Rabe, K. M.; Kaxiras, E.; Joannopoulos, J. D.
{\it Phys. Rev. B} {\bf 1990}, {\it 41}, 1227.

\bibitem{perdew81} Perdew, J. P.;  Zunger, A. {\it Phys. Rev. B} {\bf 1981}, {\it 23},
5048.

\bibitem{martin} See Eq. (17) in: Ihm, J.; Zunger, A.; Cohen, M. {\it J. Phys. C} {\bf 1979},
{\it 12}, 4409.

\bibitem{nonc} (a) Clemente-Le\'on, M.; Coronado, E.; Gal\'an-Mascar\'os, J. R.; G\'omez-Garc\'ia, C. J.;
{\it Chem. Commun.} {\bf 1997}, 1727;  (b) Coronado, E.; Gal\'an-Mascar\'os, J. R.;
G\'omez-Garc\'ia, C. J.; Ensling, J; G\"utlich, P. {\it Chem. Eur. J.} {\bf 2000} {\it 6}, 552.

\suppressfloats

\end{references}
\end{document}